\documentclass[prl,twocolumn,floats]{revtex4}

\usepackage{graphics,graphicx,epsfig}
\usepackage{amssymb}
\usepackage{epsf,epstopdf,wrapfig}
\usepackage{color}
\usepackage{natbib} 

\usepackage{natbib}

\newcommand{\beq}{\begin{equation}}
\newcommand{\eeq}{\end{equation}}
\newcommand{\beqn}{\begin{eqnarray}}
\newcommand{\eeqn}{\end{eqnarray}}
\newcommand {\e}[1]{\mathrm{~#1}}    
\newcommand {\E}[1]{\cdot 10^{#1}}		

\begin{document}

\title{Stimulus-dependent maximum entropy models of neural population codes}

\author{Einat Granot-Atedgi$^{=,a}$, Ga\v{s}per Tka\v{c}ik\footnote{$\;\;$gtkacik@ist.ac.at\\=,$\equiv$ equal contributions}$^{=,b}$, Ronen Segev$^{\equiv,c}$, and Elad Schneidman$^{\equiv,a}$}

\affiliation{$^a$Department of Neurobiology, Weizmann Institute of Science, 76100 Rehovot, Israel\\
$^b$Institute of Science and Technology Austria, Am Campus 1, A-3400 Klosterneuburg, Austria\\
$^c$Facutly of Natural Sciences, Department of Life Sciences and Zlotowski Center for Neuroscience, Ben Gurion University of the Negev, 84105 Be'er Sheva, Israel}

\date{\today}

\begin{abstract}
Neural populations encode information about their stimulus in a collective fashion, by joint activity patterns of spiking and silence. A full account of this mapping from stimulus to neural activity is given by the conditional probability distribution over  neural codewords given the sensory input. To be able to infer a model for this distribution from large-scale neural recordings, we introduce  a stimulus-dependent maximum entropy (SDME) model---a minimal extension of the canonical linear-nonlinear model of a single neuron, to a pairwise-coupled neural population.
The model is able to capture the single-cell response properties as well as the correlations in neural spiking due to shared stimulus and due to effective neuron-to-neuron connections.  Here we show that in a population of 100 retinal ganglion cells in the salamander retina responding to temporal white-noise stimuli, dependencies between cells play an important encoding role. As a result, the SDME model gives a  more accurate account of single cell responses and in particular outperforms uncoupled models in reproducing the distributions of codewords emitted in response to a stimulus. We show how the SDME model, in conjunction with static maximum entropy models of population vocabulary, can be used to estimate information-theoretic quantities like surprise and information transmission in a neural population.
\end{abstract}

\maketitle

\section{Introduction}

Neurons represent and transmit information using temporal sequences of short stereotyped bursts of electrical activity, or spikes \cite{spikes}. Much of what we know about this encoding has been learned by studying the mapping between stimuli and responses at the level of single neurons, and building detailed models of what stimulus features drive a single neuron to spike \cite{baa1,rdr, schwartz06}. In most of the nervous system, however, information is  represented by joint activity patterns of spiking and silence over populations of cells. In a sensory context, these patterns can be thought of as codewords that convey information about external stimuli to the central nervous system. One of the challenges of neuroscience is  to understand the neural \emph{codebook}---a map from the stimuli to the neural codewords---a task made difficult by the fact that neurons respond to the stimulus neither deterministically nor independently. 

The structure of correlations among the neurons determines the organization of the code, that is, how different stimuli are represented by the population activity \cite{Stopfer+al_97, Riehle+al_97,Harris+al_03, Averbeck+Lee_04}. These correlations also determine what the brain, having no access to the stimulus apart from  the spikes coming from the sensory periphery, can learn about the outside world \cite{Brunel98,Abbott99,Sompolinsky01}. The  source of these correlations, which arise either from the correlated external stimuli to the neurons, from ``shared'' local input from other neurons, or from ``private'' independent noise, have been heavily debated \cite{Schneidman+al_03a,Pola03,Nirenberg-Latham-03,Averbeck+al_06}. In many neural systems, the correlation between pairs of (even nearby or functionally similar) neurons was found to be weak \cite{bair+al_01,schneidman06,Ecker10}. Similarly, the redundancy between pairs in terms of the information they convey about their stimuli was also typically weak \cite{puchalla05,Narayanan+al_05,Chechik+al_06}. The low correlations and redundancies between pairs of neurons therefore led to the suggestion that neurons in larger populations might encode information independently \cite{nirenberg}, which was echoed by theoretical ideas of maximally efficient neural codes \cite{barlow,attick+redlich_90,Barlow_01}. 

Recent studies of the neural code in large populations have, however, revealed that while the typical pairwise correlations may be weak, larger populations of neurons can nevertheless be strongly correlated as a whole \cite{Schnitzer+Meister_03,schneidman06,preprint,shlens+al_06,tangetal,shlens+al_09,marre,preprint2,Ganmor+al_11a}. Maximum entropy models of neural populations have shown that such strong network correlations can be the result of  collective effects of  pairwise dependencies between cells, and, in some cases, of sparse high-order dependencies \cite{schneidman06,jdhigh,ganmor}. Most of these studies have characterized the strength of  network effects and spiking synchrony at the level of the total \emph{vocabulary} of the population, i.e. the distribution of codewords averaged over all the stimuli. It is not immediately clear how these findings affect stimulus encoding, where one needs to distinguish the impact of correlated stimuli that the cells receive (``stimulus correlations''), from the impact of  co-variance of the cells conditional on the stimulus (``noise correlations''). For small populations of neurons, it has been shown that taking into account correlations for decoding or reconstructing the stimulus can be beneficial compared to the case where correlations are neglected (e.g.~\cite{warland97,Dan+al_98,Hatsopoulos+al_98,Brown+al_98,ganmor}). Similarly, generalized linear models highlighted the importance of dependencies between cells in accounting for correlations between pairs and triplets of retinal ganglion cell responses \cite{pillow08}. 

Here we present a new encoding model that allows us to study in fine detail the  codebook of large neural populations. Importantly, this model gives a joint probability distribution over the activity patterns of the whole population for a given stimulus, while capturing both the stimulus and noise correlations. This new model belongs to a class of maximum entropy models with strong links to statistical physics \cite{still,preprint,roudi1,roudi2,roudi3,sessak,monasson,monasson2,macke,cessac,thierry,roudi4,cessac2} and is directly related to maximum entropy models of neural vocabulary \cite{schneidman06,preprint,shlens+al_06,tangetal,marre,shlens+al_09,preprint2}, allowing us estimate the entropy and its derivative quantities for the neural code. In sum, the maximum entropy framework enables us to progress towards our goal of focusing  attention on the level of joint patterns of activity, rather than capturing low-level statistics (e.g., the individual firing rates) of the neural code alone.

We start by showing that linear-nonlinear (LN) models of retinal ganglion cells responding to spatially unstructured stimuli capture a significant part of the single neuron response, but still miss much of the  detail;  in particular, we show that they fail to capture the correlation structure of firing among the cells. We next present our new \emph{stimulus-dependent maximum entropy} (SDME) model, which is a hybrid between linear-nonlinear models for single cells and the pairwise maximum entropy models. Applied to groups of $\sim 100$ neurons recorded simultaneously, we find that SDME models outperform the LN models for the stimulus-response mapping of single cells and, crucially, give a significantly better account of the distribution of  codewords in the neural population.

\section{Results}
 \begin{figure}[] 
\centering
\includegraphics[width=3.5in]{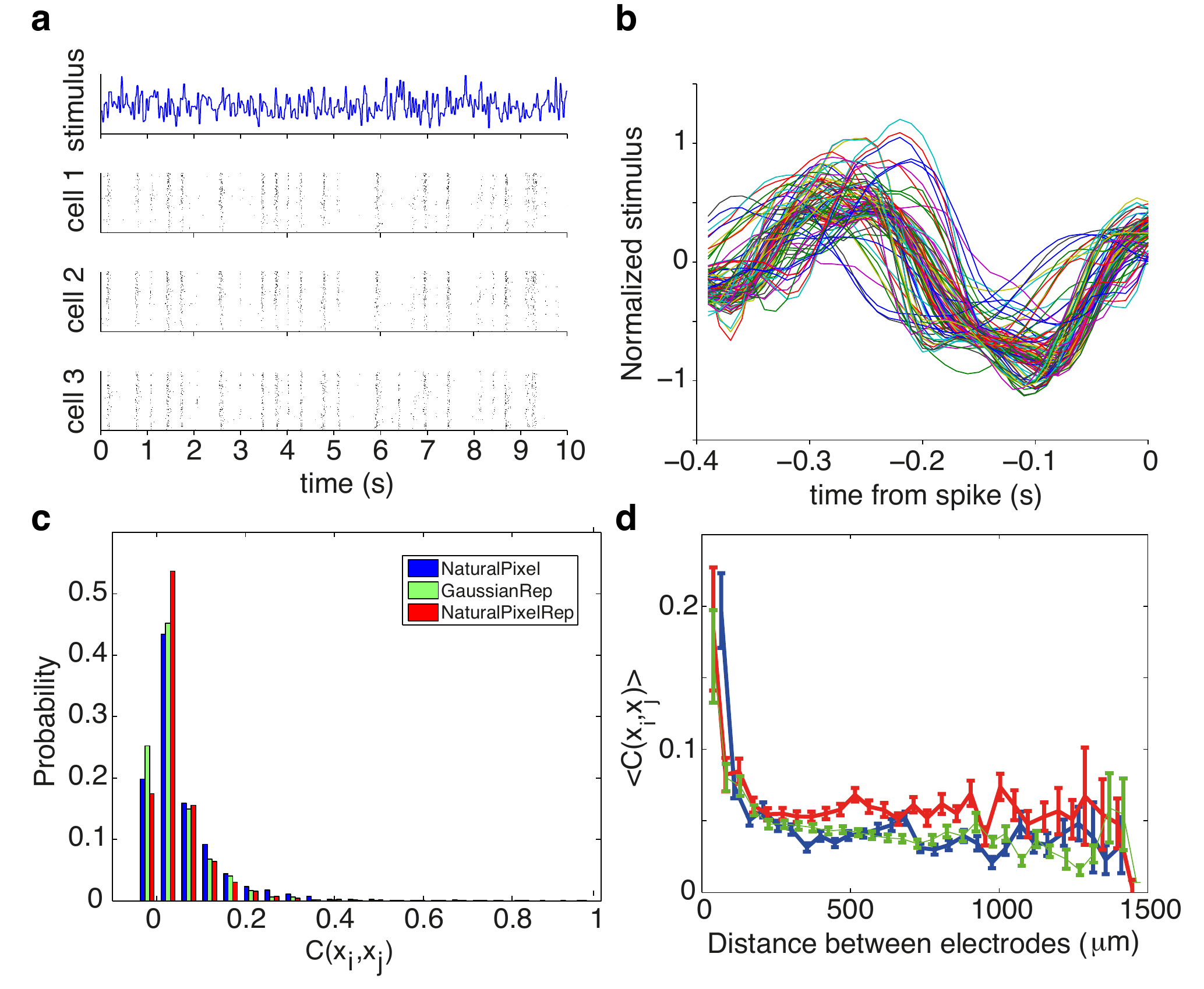} 
\caption{{\bf Response of a large population of ganglion cells to a 10 s long repeated visual stimulus}. {\bf (a)} White noise uncorrelated Gaussian stimulus presented at $30\e{Hz}$ and the spiking patterns of 3 cells to repeated presentations of the stimulus. {\bf (b)} Spike-trigerred averages of 110 simultaneously recorded cells; a subset of 100 cells was chosen for further analysis. {\bf (c)} The histogram of pairwise correlations between cells for repeated Gaussian white noise stimulus (green), repeated natural pixel movie (red), and non-repeated natural pixel movie (blue) \cite{ganmor}. {\bf (d)} Average pairwise correlation coefficient between cells as a function of the distance (mean and std  are across pairs of cells at a given distance).}
\label{f1}
\end{figure}

We recorded the simultaneous spiking activity of $\sim 110$ ganglion cells from the salamander retina \cite{segev04}, presented  with repeats of a $10\e{s}$ long full-field flicker (``Gaussian FFF'') movie, where the light intensity on the screen was sampled independently from a Gaussian distribution with a frequency of $30\e{Hz}$ (Fig.~\ref{f1}a). This ``frozen noise'' stimulus was repeated 726 times, for a total of $\sim 2\e{h}$ of stimulation. Most of the recorded cells exhibited temporal OFF-like behaviors (Fig.~\ref{f1}b). We chose for further analysis $N=100$ cells that were reliably sorted, demonstrated a robust and stable response over repeats, and generated at least $2500$ spikes during the course of the experiment. 

\begin{figure}[] 
\centering
\includegraphics[width=3.5in]{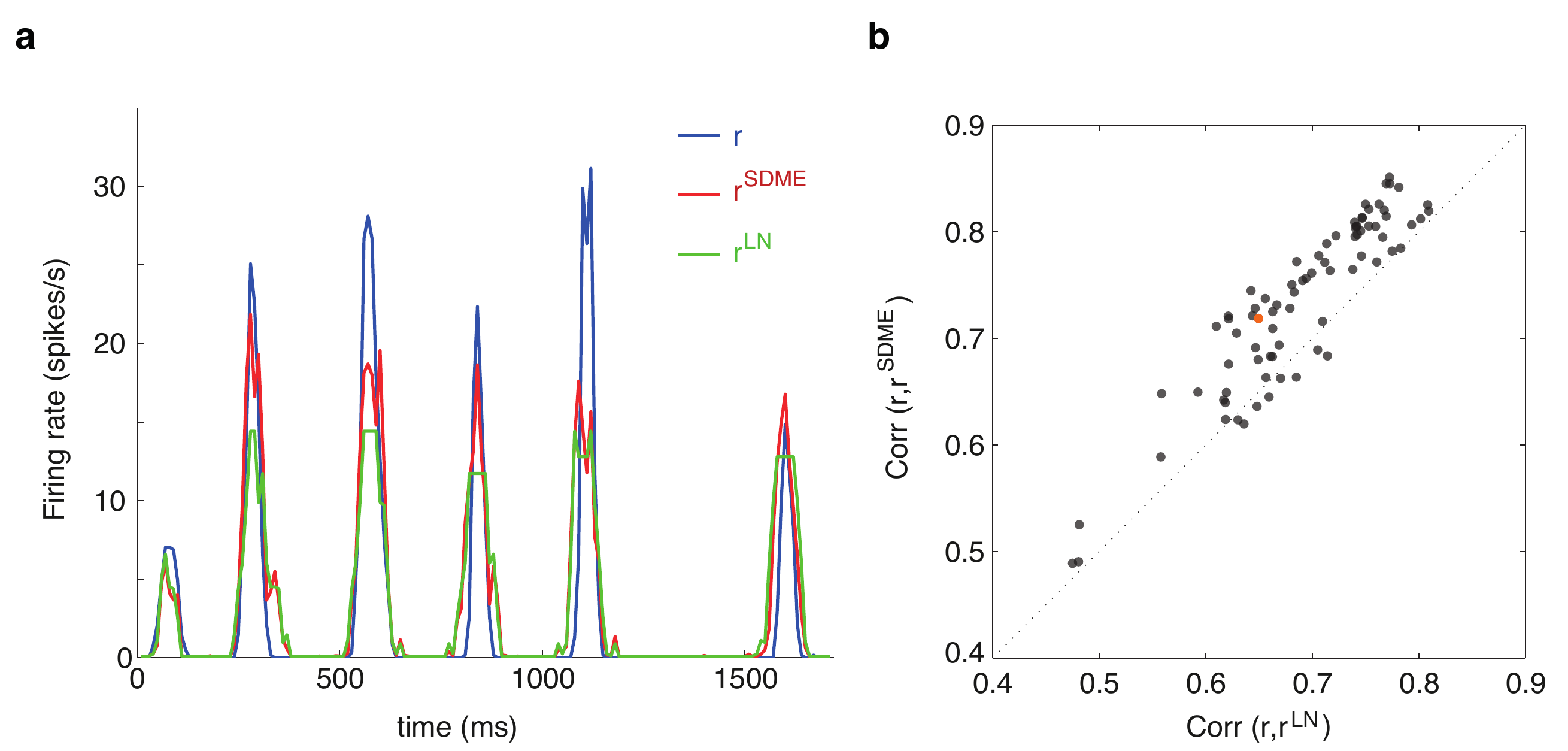} 
\caption{{\bf SDME model predicts the firing rate of single cells better than LN models.} {\bf (a)} Example of the PSTH segment for one cell (blue),  the best prediction of an LN model (green) and of a SDME model (red). {\bf (b)} Correlation between the true PSTH and SDME model prediction (vertical axis) vs. the correlation between the true PSTH and the LN model prediction (horizontal axis); each plot symbol is a separate cell, dotted line shows equality. The neuron chosen in panel (a) is shown in orange.}
\label{f2}
\end{figure}

We discretized neural responses into $\Delta t=10\e{ms}$ bins, and represented the activity of the neurons in response to the stimulus as binary words in each of the time bins. If neuron $i=1,\dots,N$ was active in time bin {\em t}, we denoted a spike (or more spikes) as $x_i(t) =1$, and $x_i(t) =0$ if it was silent. In this representation, the whole experiment yielded a total of about $T\sim 7.3\E{5}$ binary word samples. Using repeated presentations of the same movie, we estimated the average response of each of the cells across repeats, $r_i(t) = \langle x_i(t)\rangle_{rep}$, or the peri-stimulus time histogram (PSTH). Following Refs.~\cite{fairhall_berry,schwartz06}, we fitted a linear-nonlinear model for each of the cells  in the experiment, such that the predicted rate $r^{LN}_i(t)=\mathcal{N}_i(\mathbf{k}_i\cdot\mathbf{s}(t))$, where $\mathbf{k}_i$ is a linear filter matched for the $i$-th cell, $\mathcal{N}_i$ is its point-wise nonlinear function, and $\mathbf{s}(t)$ is the stimulus fragment from time $t-\tau$ until $t$ (here we used $\tau=400\e{ms}$, making $\mathbf{s}(t)$ a vector of light intensities with 40 components). Linear filters were reconstructed using reverse correlation (spike-triggered average), and nonlinearities were obtained by histograming $P(\mathbf{k}_i\cdot \mathbf{s}(t)|\mathrm{spike})$ into $K=20$ adaptively-sized bins and obtaining  $r^{LN}_i(t)=\mathcal{N}_i(\mathbf{k}_i\cdot \mathbf{s})=P(\mathrm{spike}|\mathbf{k}_i\cdot \mathbf{s}(t))$ by inverting $P(\mathbf{k}_i\cdot \mathbf{s}(t)|\mathrm{spike})$ using  Bayes' rule. These LN models captured most of structure of the PSTH, yet as the example cell in Fig.~\ref{f2}a shows, they often misestimated the exact firing rates of the neuron, or sometimes even missed parts of the neural response altogether. In the Gaussian FFF condition, the normalized (Pearson) correlation between the measured and predicted PSTH, $\mathrm{Corr}(r_i(t),r^{LN}_i(t))$, was $0.69\pm 0.06$ (mean $\pm$ std across 100 cells).

The performance gap of the canonical LN models in predicting single neuron responses suggests that either the single-neuron models need to be improved to account for the observed behavior, or that interactions between neurons play an important encoding role and need to be included. While firing rate prediction performance can be improved for single neurons by models with higher-dimensional stimulus sensitivity (e.g. \cite{fairhall_berry,tkacik_hos}) or dynamical aspects  of spiking behavior (e.g. \cite{keat,ozuysal}), previous work, as well as the results below,  demonstrated that even conditionally-independent models which by construction perfectly reproduce the firing rate behavior of single cells, often fail to capture the measured correlation structure of firing between pairs of cells, as well as higher-order statistical structure \cite{schneidman06}.
 
We find  two salient features of the correlations between pairs of neurons: (i) the pairwise correlations between cells in response to the Gaussian FFF movie are typically weak but are not zero (Fig.~\ref{f1}c, consistently with previous reports \cite{schneidman06,preprint,preprint2});  (ii) the correlation in neural activities shows a fast decay with distance despite the infinite correlation length of the stimulus, but the decay does not reach zero correlation  even at relatively large distances (Fig.~\ref{f1}d). This salient structure, along with any other potential statistical correlation at the pairwise order, is  characterized by the covariance matrix of activities, $C_{ij}=\langle x_ix_j\rangle-\langle x_i\rangle\langle x_j\rangle$, where the averages are taken across time and repeats. 

We would like to find a model of the neural code that will be able to reproduce these pairwise statistics. 
Motivated by the shortcomings of the canonical LN model, we therefore asked  whether a model that  combined the LN (receptive-field based) aspect of single cells with the interactions between cells, could give a better account of the neural stimulus-response mapping. Importantly, the new model should capture not only the firing rate of single cells 
and  the full pairwise correlation structure between them, but should also accurately predict the full distribution of the joint activity patterns across the whole population. Because the joint distributions of activity are high-dimensional (e.g., the distribution over codewords across the duration of the experiment, $P(\{x_i\})$, has $2^N$ components), this is a very demanding benchmark for any model.

Here we propose the simplest extension to the conditionally-independent set of LN models for each cell in the recorded population, by including pairwise couplings between cells, so that the spiking of cell $i$ can increase or decrease the probability of spiking for cell $j$ \cite{tkacik_phd,granot}. In contrast to previous proposals, this coupling will be introduced so that the resulting model is a maximum-entropy model for $P(\{x_i\}|\mathbf{s})$, the conditional distribution over population activity patterns given the stimulus. We recall that the maximum entropy models give the most parsimonious probabilistic description of the joint activity patterns, which perfectly reproduces a  chosen set of  measured statistics over these patterns, without making any additional assumptions \cite{jaynes_57}.

We start by introducing the least structured (maximum entropy) distribution $P(x_1,x_2,\dots,x_N|t)$  that reproduces exactly the observed average firing rate for each time bin $t$ and for each neuron $i$, $r_i(t)=\langle x_i(t)\rangle_{data}=\langle x_i(t) \rangle_P$, as well as the overall correlation matrix $C_{ij}$ between all pairs of cells (c.f.~\cite{tkacik10}). Thus, we seek $P(\{x_i\}|t)$ that  maximizes $\mathcal{L}$:
\begin{eqnarray}
\mathcal{L}\left[P(\{x_i\}|t) \right]& = & -\sum_{\{x_i\},t}  P(\{x_i\}|t)  \log_2 P(\{x_i\}|t) \nonumber \\
& + & \sum_{i,t} \alpha_i(t) [\langle x_i(t) \rangle_{P} - \langle x_i(t) \rangle_{data}] \nonumber \\
&+ & \frac{1}{2} \sum_{ij} \beta_{ij} [\langle x_i x_j \rangle_{P,t} - \langle x_i x_j \rangle_{data}] \nonumber \\
&+& \sum_{\{x_i\},t}  \mu(t)[ P(\{x_i\}|t) - 1], \label{eqtl}
\end{eqnarray}
where the subscript to brackets $\langle \cdot \rangle$ denotes whether the averaging is done over the maximum entropy distribution ($P$), or over the recorded data; Lagrange multipliers $\mu$ ensure that the distributions are normalized. This is an optimization problem for parameters $\alpha_i(t)$ and $\beta_{ij}$, which has a unique solution  since the entropy is convex. The functional form of the solution to this optimization problem is well-known; in our case it can be written as
\begin{eqnarray}
&&P^{TDME}(\{x_i\}|t) =  \label{eqt}  \\
&&\frac{1}{Z(t)} \exp \left( \sum_{i=1}^N \alpha_i(t) x_i + \frac{1}{2} \sum_{i,j=1}^N \beta_{ij} x_i x_j \right), \nonumber
\end{eqnarray}
where the individual time-dependent parameters for each of the cells, $\alpha_i(t)$, and the stimulus-independent pairwise interaction terms $\beta_{ij}$,  are set to match the measured firing rates $r_i(t)$ and the pairwise correlations $C_{ij}$;  $Z(t)$ is a normalization factor or partition function for each time bin $t$, given by $Z(t)=\sum_{\{x_i\}} \exp \left( \sum_i \alpha_i(t) x_i + \frac{1}{2} \sum_{ij} \beta_{ij} x_i x_j \right)$. 

The time-dependent maximum entropy (TDME) model in Eq.~(\ref{eqt}) is equivalent to an Ising model from physics, where the single-cell parameters are time-dependent local fields acting on each of the neurons (spins), and static (stimulus-independent) infinite-range interaction terms couple each pair of spins. In the limit where interactions go to zero, $\beta_{ij}=0$, the model in Eq.~(\ref{eqt}) becomes the full conditionally-independent model, itself a maximum entropy model that reproduces exactly the firing rate of every neuron, $r_i(t)$; in this case the probability distribution factorizes, and the solution for $\alpha_i(t)$ and $Z(t)$ becomes trivially computable from the firing rates, $r_i(t)$. Time-dependent maximum entropy models are powerful, since they make no assumptions about how the stimulus drives the response; they can serve as useful benchmarks for other models (especially the conditionally independent model with $\beta_{ij}=0$). On the other hand, these models require repeated stimulus presentations to fit, involve a number of parameters that grows linearly with the duration of the stimulus, do not generalize to new stimuli, and do not provide an explicit map from  the stimuli to the responses.

To make a direct link to the stimulus and allow comparison with a set of uncoupled LN models, we take the general time-dependent maximum entropy model  of Eq.~(\ref{eqt}) and specialize it to a particular form of stimulus dependence. Rather than having an arbitrary time-dependent parameter for every neuron for each time bin, $\alpha_i(t)$, we assume that this dependence takes place through the stimulus projection alone, i.e. $\alpha_i(t)=\alpha_i(\mathbf{k}_i\cdot \mathbf{s}(t))$, much like in an LN model, where the neural firing depends on the value of the stimulus projection onto the linear filter $\mathbf{k}_i$. This choice is made purely for the  sake of convenience: the model could be generalized to, e.g., neurons that depend on two linear projections of the stimulus, by making $\alpha_i$ depend jointly on $(\mathbf{k}_1\cdot \mathbf{s}(t),\mathbf{k}_2\cdot\mathbf{s}(t))$, although such models would be progressively more difficult to infer from data.

Concretely, we estimated the linear filter $\mathbf{k}_i$ for each cell $i$ using reverse correlation, and convolved the filter with the stimulus sequence, $s(t)$, to get the ``generator signal'' $g_i(t)=\mathbf{k}_i\cdot \mathbf{s}(t)$. We then looked for the maximum entropy probability distribution $P(\{x_i\}|\mathbf{s}(t))$, by requiring that the average firing rate of  every cell given the generator signal is the same in the data and under the model, i.e. $\langle x_i ( g_i) \rangle_{data} =  \langle x_i (g_i)  \rangle_{P} $ (see Methods); as before, we also required that the model reproduced the overall correlation between every pair of cells, $C_{ij}$.  This gives then a stimulus-dependent maximum entropy (SDME) model, which takes the following form:  
\begin{eqnarray}
&&P^{SDME}(\{x_i\}|\mathbf{s}(t)) =\label{eqs}  \\   
&&\frac{1}{Z(\mathbf{s}(t))} \exp \left( \sum_{i=1}^N \alpha_i(g_i(t)) x_i + \frac{1}{2} \sum_{i,j=1}^N \beta_{ij} x_i x_j \right). \nonumber
\end{eqnarray}
The parameters of this model are:  $N\times (N-1)/2$ couplings $\beta_{ij}$, $K\times N$ parameters $\alpha_i$, and a linear filter $\mathbf{k}_i$ for each cell. We used a Monte Carlo based gradient descent learning procedure to find the model parameters $\alpha,\beta$ numerically (see Methods). 

By construction, the SDME model exactly reproduces the covariance in activities, $C_{ij}$, between all pairs of cells, and also the LN model properties of every cell: an arbitrary nonlinear function $\mathcal{N}$ can be encoded by properly choosing how parameters $\alpha_i$ depend on the linear projections of the stimulus, $g_i$. We can construct a maximum entropy model with $\beta_{ij}=0$ (no constraints on the pairwise correlations $C_{ij}$). The result is a set of uncoupled (conditionally independent) LN models: 
\begin{eqnarray}
P^{LN}(\{x_i\}|\mathbf{s}(t)) &\equiv& \prod_{i=1}^N \frac{1}{\tilde{Z}_i(\mathbf{s}(t))} \exp \left( \tilde{\alpha}_i(g_i(t)) x_i \right) \label{eqs0}\\
&=&\prod_{i=1}^N\mathcal{N}_i(\mathbf{k}_i\cdot\mathbf{s}(t)). \nonumber
\end{eqnarray}
In sum, the time-dependent maximum entropy (TDME) model of Eq.~(\ref{eqt}) is an extension of conditionally independent model that additionally reproduces the measured pairwise correlations between cells. In a directly analogous way, the stimulus-dependent maximum entropy (SDME) model of Eq.~(\ref{eqs}) is an extension to the set of uncoupled LN models, Eq.~(\ref{eqs0}), that additionally reproduces the measured pairwise correlations between cells. Because $P^{SDME}$ (Eq.~\ref{eqs}) agrees with $P^{LN}$ (Eq.~\ref{eqs0})  exactly in all constrained single-neuron statistics,  any improvement in prediction of the SDME, be it in the firing rate or the codeword distributions, can be directly ascribed to the effect of the interaction terms, $\beta_{ij}$. 

\begin{figure*}[] 
\centering
\includegraphics[width=7in]{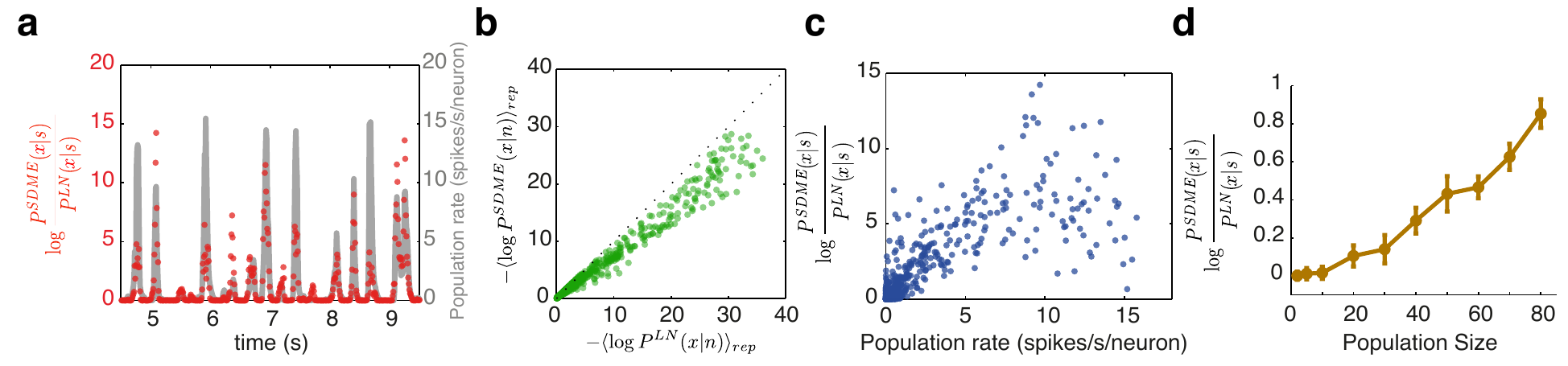} 
\caption{{\bf SDME model predicts population activity patterns for $N=100$ neurons  better than conditionally independent LN models.} {\bf (a)} The log-likelihood ratio of the population firing patterns under the SDME model and under a collection of LN models, shown as a function of time (red) for an example stimulus repeat. For reference, the average population firing rate is shown in grey. 
{\bf (b)} A scatterplot of the log-likelihoods under the SDME and LN models; each dot represents an average over activity patterns $\{x_i\}$ (across repeats) at a given time bin; dotted line shows equality. {\bf (c)} The log-likelihood ratio under the SDME and LN models as a function of the average firing rate in the population; SDME outperforms LN models in particular for patterns with more spiking activity. {\bf (d)} The average likelihood ratio between the SDME and LN models as a function of the population size, $N$ (error bars = std over 10 randomly chosen groups of neurons at that $N$). }
\label{f3}
\end{figure*}

We found that the SDME  predicted the firing rate of single cells better than the LN models, with the normalized correlation coefficient between the true and predicted firing rate, $\mathrm{Corr}(r_i(t),r_i^{SDME}(t))$ being $0.74\pm 0.06$ (mean $\pm$ std across 100 cells), as shown in Fig.~\ref{f2}b. 
The differences between the SDME and the  LN models become  more striking on the level of the activity patterns of the whole population. Figures~\ref{f3}a,b show the log-likelihood ratio for the population activity patterns $\mathbf{x}=\{x_i\}$ under the two models, showing that the SDME can be orders of magnitude better in predicting the population neural response. These differences are large in particular for those stimuli that elicit a strong response (Fig.~\ref{f3}c), that is, precisely where the response consists of synchronous spiking and the structure of the codewords can be nontrivial. Moreover, the difference between the models becomes increasingly significant with the size of the population $N$, and particularly apparent for groups of more than 20 cells  (Fig.~\ref{f3}d).

We next examined how well various models for the neural codebook, $P(\{x_i\}|\mathbf{s})$, explain the total  vocabulary, that is, the distribution of neural codewords observed across the whole duration of the experiment, $P(\{x_i\})=\langle P(\{x_i\}|\mathbf{s}(t))\rangle_t$. Despite the nominally large space of possible codewords---much larger than the total number of samples in the experiment ($2^N\gg T$)---the sparsity of spikes and the correlations between neurons restrict the vocabulary to a much smaller set of patterns. Some of these occur many times during our stimulus presentation, allowing us to estimate their empirical probability, $P^{data}(\{x_i\})$, directly from the experiment, and compare it to the model prediction \cite{ganmor}. The most prominent example of such frequently observed codewords is the silent pattern, $x_i=0$, which is seen $\sim 72\%$ of the time. 

\begin{figure}[tb] 
\centering
\includegraphics[width=3.5in]{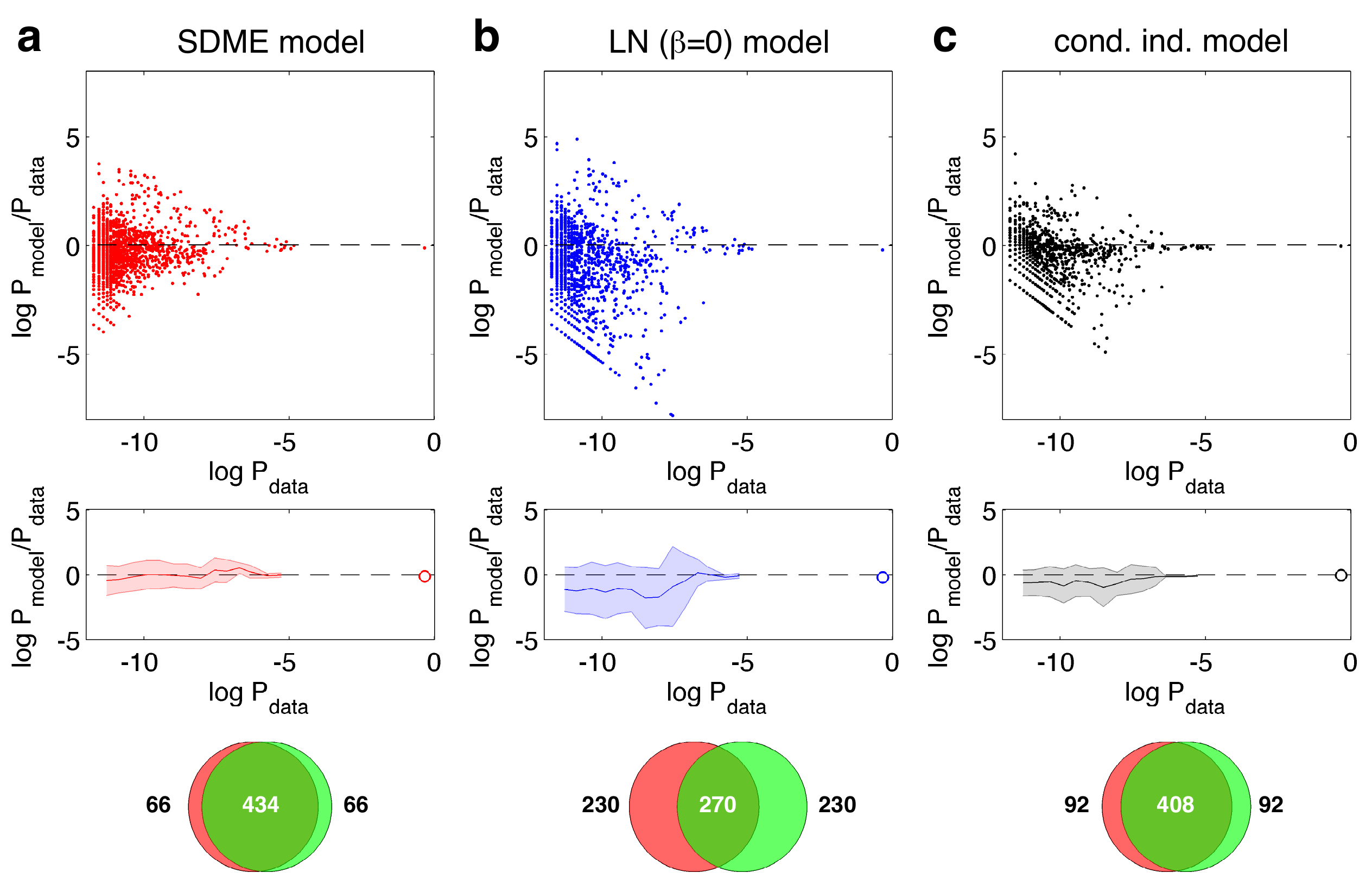} 
\caption{{\bf The performance of various models in accounting for the total vocabulary of the population, $P(\{x_i\})$.} The results for the SDME model are shown in {\bf (a)}, the results for an uncoupled set of LN models in {\bf (b)}, the results for a full conditionally independent model in {\bf (c)}.  The first row displays the log ratio of model to empirical probabilities for various codewords (dots), as a function of that codeword's empirical frequency in the recorded data. The model probabilities were estimated by generating Monte Carlo samples from the corresponding model distributions (see Methods); only patterns that were generated in the MC run as well as found in the recorded data are shown. The second row summarizes this scatterplot by binning codewords according to their frequency, and showing the average log probability ratio in the bin (solid line), as well as the $1$ std scatter across the codewords in the bin (shaded area). The highly probable all-silent state, $\{x_i\}=0$, is shown separately as a circle. The third row shows the overlap between 500 most frequent patterns in the data and 500 most likely patterns generated by the model (see text). } 
\label{f5}
\end{figure}

Figure~\ref{f5} shows the likelihood ratio of the model probability and empirical probability for various codewords observed in the experiment, as a function of the rate at which these codewords appear in the experiment.  While SDME model in Fig.~\ref{f5}a does not predict the frequencies of all patterns perfectly, it strongly outperforms the uncoupled set of LN models in Fig.~\ref{f5}b, and has a slightly better performance than the full conditionally independent model (Fig.~\ref{f5}c), despite the fact that the latter is determined by $N\times 1000=1\E{5}$ parameters, the firing rates of every cell in every time bin. On average, SDME predicts the probabilities of the patterns of activity with no bias, and with a standard deviation of $\log(P^{SDME}/P^{data})$ of about 1; uncoupled LN models in comparison are biased and have a spread that is more than twice as large. Even more striking is the fact that LN models assign such low probabilities to some codewords that they are never generated during our Monte Carlo sampling (and are therefore not even shown in scatterplots of Fig.~\ref{f5}), while they are frequently observed in the experiment. This discrepancy is quantified by enumerating the $M$ most probable patterns in the data and in the model (by sampling; see Methods), and  measuring the size of the intersection of the two sets of patterns; in other words, we ask if  the model is even able to access all the patterns that one is likely to record in the experiment. As shown in the third row of Fig.~\ref{f5}, SDME does well on this task, with 434 codewords in the intersection of the 500 most likely patterns in the data and the model; this is a much better performance than for the uncoupled model, and slightly better than for the conditionally independent model.

\begin{figure}[tb] 
\centering
\includegraphics[width=3in]{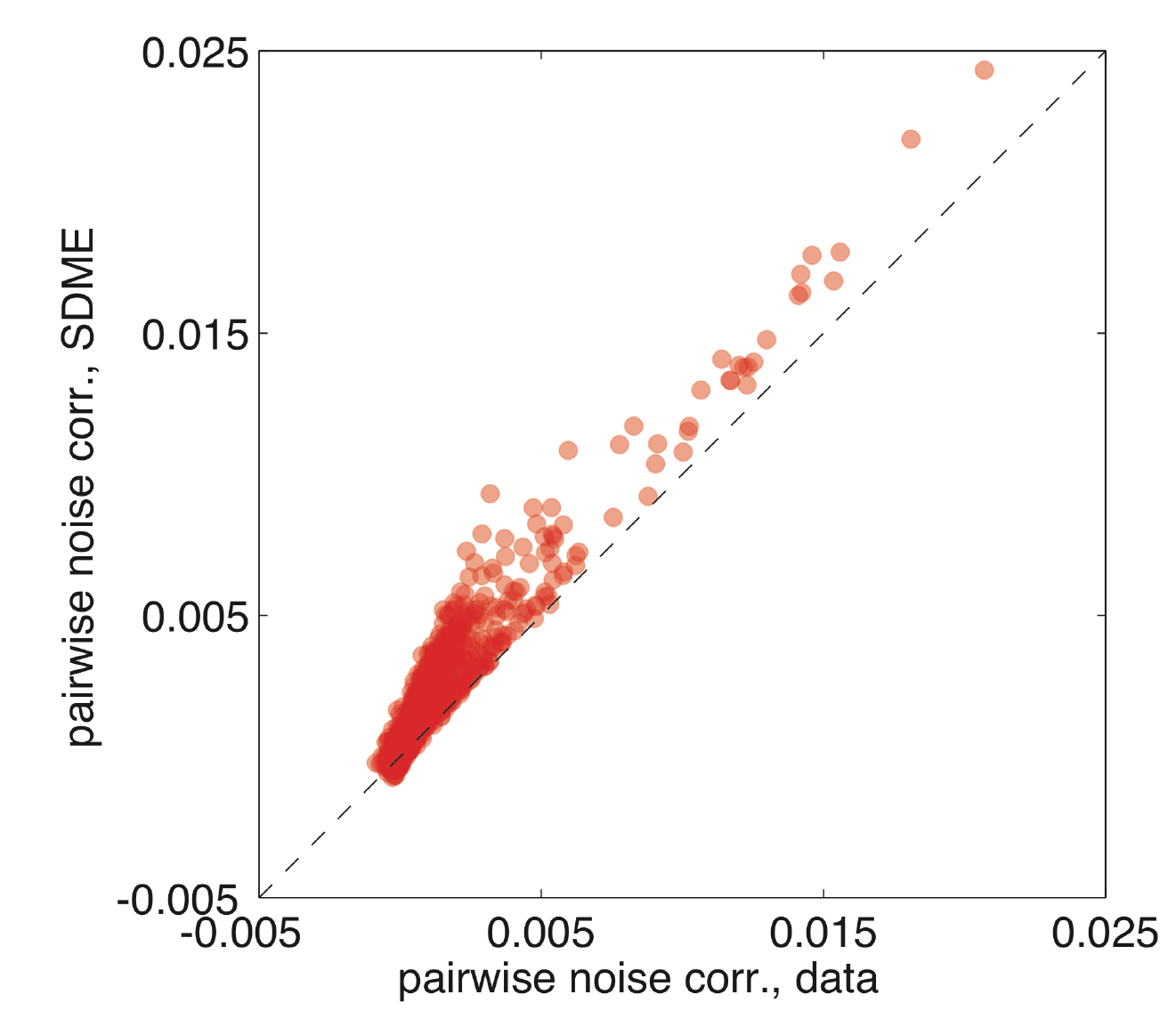} 
\caption{{\bf Measured vs predicted noise correlations for the SDME model.} Noise correlation (see text) is estimated from recorded data for every pair of neurons, and plotted against the noise correlation predicted by the SDME model (each pair of neurons = one dot; shown are $N(N-1)/ 2$ dots  for $N=100$ neurons). Conditionally independent models predict zero noise correlation for all pairs. }
\label{f6}
\end{figure}

The SDME model was constructed to capture exactly the total correlations in neuronal spiking, $C_{ij}=\langle x_i x_j\rangle - \langle x_i\rangle\langle x_j\rangle$. With repeated stimulus, this total correlation can be broken down into the signal and noise components. The signal correlations, $C_{ij}^s$, are inferred by applying the same formula as for the total correlation, but on the spiking raster where the repeated trial indices have been randomly and independently permuted for each time bin. This removes any correlation due to interactions between spikes on simultaneously recorded trials, and only leaves the correlations induced by the response being locked to the stimulus. The noise correlation, $C^n_{ij}$, is then defined as the difference between the total and the signal components, $C^n_{ij} = C_{ij}-C_{ij}^s$. We calculated the noise correlations between all pairs in our $N=100$ neuron dataset. By their definition,  the conditionally independent models cannot reproduce $C^n_{ij}$, which are always zero. To assess the performance of the SDME, we drew samples from our model distribution using the Monte Carlo simulation and compared the noise correlations in the simulated rasters to the true noise correlations. The model prediction tightly correlates with the measured values, as  shown in Fig.~\ref{f6}. We observe a systematic deviation of $\sim 25\%$, most likely because the assumed dependence on the stimulus through one linear filter per neuron is insufficient to capture the complete dependence on stimulus, thereby underestimating the full structure of stimulus correlation and inducing an excess in the noise correlation. Despite this, the  degree of correspondence in noise correlations observed in Fig.~\ref{f6}  is telling us that SDME has clearly captured a large amount of noise covariance structure in neural firing.

\begin{figure}[tb] 
\centering
\includegraphics[width=3.5in]{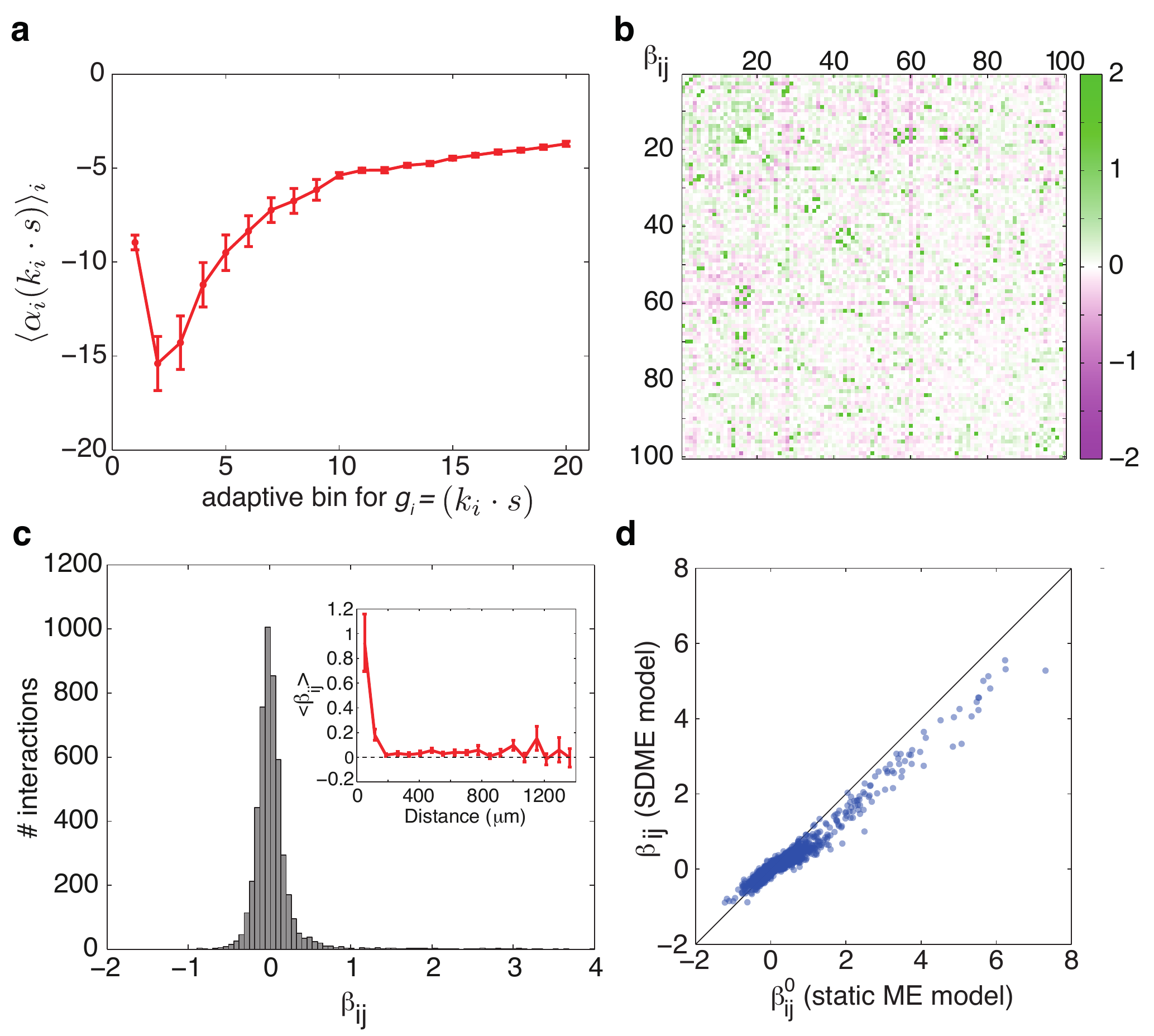} 
\caption{{\bf SDME model parameters.} {\bf (a)} Average values of the LN-like driving term, $\alpha_i(g_i)$, where $g_i=\mathbf{k}_i\cdot\mathbf{s}$, across all cells $i$ (error bars = std across cells), for each of the $K=20$ adaptive bins for $g_i$ (see Methods). {\bf (b)} Pairwise interaction map $\beta_{ij}$ of the SDME model, between all $N=100$ neurons in the experiment. {\bf (c)} Histogram of pairwise interaction values from (b), and their average value as a function of the distance between cells (inset).  {\bf (d)} For each pair of cells $i$ and $j$, we plot the value of $\beta_{ij}^0$ under the static maximum entropy model of Eq.~(\ref{me}) vs. the $\beta_{ij}$  from the stimulus-dependent maximum entropy model of Eq.~(\ref{eqs}). }
\label{f7}
\end{figure}

How should we interpret the inferred parameters of the SDME model? LN models have a clear ``mechanistic'' interpretation in terms of the cell's receptive field and the nonlinear spiking mechanism. Here, similarly, the stimulus dependent part of the model for each cell, $\alpha_i$, is a nonlinear function of a filtered version of the stimulus $g_i(t) = \mathbf{k}_i\cdot \mathbf{s}(t)$; in the absence of neuron-to-neuron couplings, the nonlinearity of every neuron would  correspond to $\mathcal{N}_i(g_i)\sim f(\alpha_i(g_i))$, where $f(\cdot)=\exp(\cdot)/(1+\exp(\cdot))$, according to Eq.~(\ref{eqs0}). The dependence of $\alpha_i$ on the stimulus projection $g_i$ is  similar across the recorded cells as shown in Fig.~\ref{f7}a; as expected, higher overlaps with the linear filter induce higher probability of spiking. 

The pairwise interaction terms in the model, $\beta_{ij}$, are symmetric, static, and stimulus independent by construction. As such, they represent only functional and not physical (i.e. synaptic) connections between the cells. Fig.~\ref{f7}b shows the pairwise interaction map for 100 cells; the histogram of their values (in Fig.~\ref{f7}c) reflects that they can be of both signs, but the distribution has a stronger positive tail, i.e. a number of cell pairs tend to spike together or be silent together with a probability than is higher than expected from their respective LN models. We can compare these interactions to the interactions of a static (non-stimulus-dependent) maximum entropy model for the population vocabulary \cite{schneidman06,shlens+al_06}:
\begin{equation} 
P^{ME}(\{x_i\}) = \frac{1}{Z_0} \exp \left( \sum_i \alpha_i^0 x_i + \frac{1}{2} \sum_{ij} \beta_{ij}^0 x_i x_j \right). \label{me}
\end{equation}
In this model for the total distribution of codewords, there is no stimulus dependence, and the parameters $\alpha_i^0$ and $\beta_{ij}^0$ are chosen to that the distribution is as random as possible, while reproducing exactly the measured mean firing rate of every neuron $\langle x_i\rangle_{data}=\langle x_i\rangle_{P^{ME}}$, and every pairwise correlation, $\langle x_ix_j\rangle_{data}=\langle x_ix_j\rangle_{P^{ME}}$, across the whole duration of the experiment.

Interestingly, we find that the pairwise interaction terms in the SDME model of Eq.~(\ref{eqs}) are closely related to the interactions in the static pairwise maximum entropy model of Eq.~(\ref{me}): SDME interactions, $\beta_{ij}$, tend to be smaller in magnitude, but have an equal sign and relative ordering, as the static ME interactions, $\beta_{ij}^0$. Some degree of correspondence is expected: an interaction between neurons $i$ and $j$ in the static ME model captures the combined effect of the stimulus and noise correlations, while in the corresponding SDME interaction, (most of) the stimulus correlation has been factored out into the correlated dynamics of the inputs to the neurons $i$ and $j$, i.e. $\alpha_i(g_i(t))$ and $\alpha_j(g_j(t))$.  The surprisingly high degree of correspondence, however, indicates that even the interactions learned from static maximum entropy models can account for, up to a scaling factor, the pairwise neuron dependencies that are \emph{not} due to the correlated stimulus inputs.

The SDME model is an approximation to the neural codebook, $P(\{x_i\}|\mathbf{s})$, while the static ME model describes the population vocabulary, $P(\{x_i\})$. With these two distributions in hand, we can explore how the  population jointly encodes the information about the stimulus into neural codewords---the joint activity patterns of spiking and silence. We make use of the fact that we can estimate the entropy of the maximum entropy distributions using a procedure of heat capacity integration, as explained in Refs.~\cite{preprint,preprint2} (see Methods). Information (in bits) per codeword is 
\begin{eqnarray}
I(\{x_i\};\mathbf{s})&=&\int d\mathbf{s}\; P(\mathbf{s}) \sum_{\{x_i\}}P(\{x_i\}|\mathbf{s})\log_2\frac{P(\{x_i\}|\mathbf{s})}{P(\{x_i\})} \nonumber \\
&=& S[P(\{x_i\})] - \langle S[P(\{x_i\}|\mathbf{s})] \rangle_{P(\mathbf{s})}; \label{itrans}
\end{eqnarray}
that is, the information can be written as a difference of the entropy of the neural vocabulary, and the noise entropy (the average of the entropy of the codebook), where the entropy is $S[p(x)]=-\int dx\; p(x) \log_2 p(x)$. Because of the maximum entropy property of our model for $P^{ME}(\{x_i\})$, the entropy of our static pairwise model in Eq.~(\ref{me}) is an upper bound on the transmitted information; expressed as an entropy rate, this amounts to $s \equiv S[P^{ME}(\{x_i\})]/\Delta t\approx 730 \e{bit/s}$. 

\begin{figure}[!t]
\centering
\includegraphics[width=3.5in]{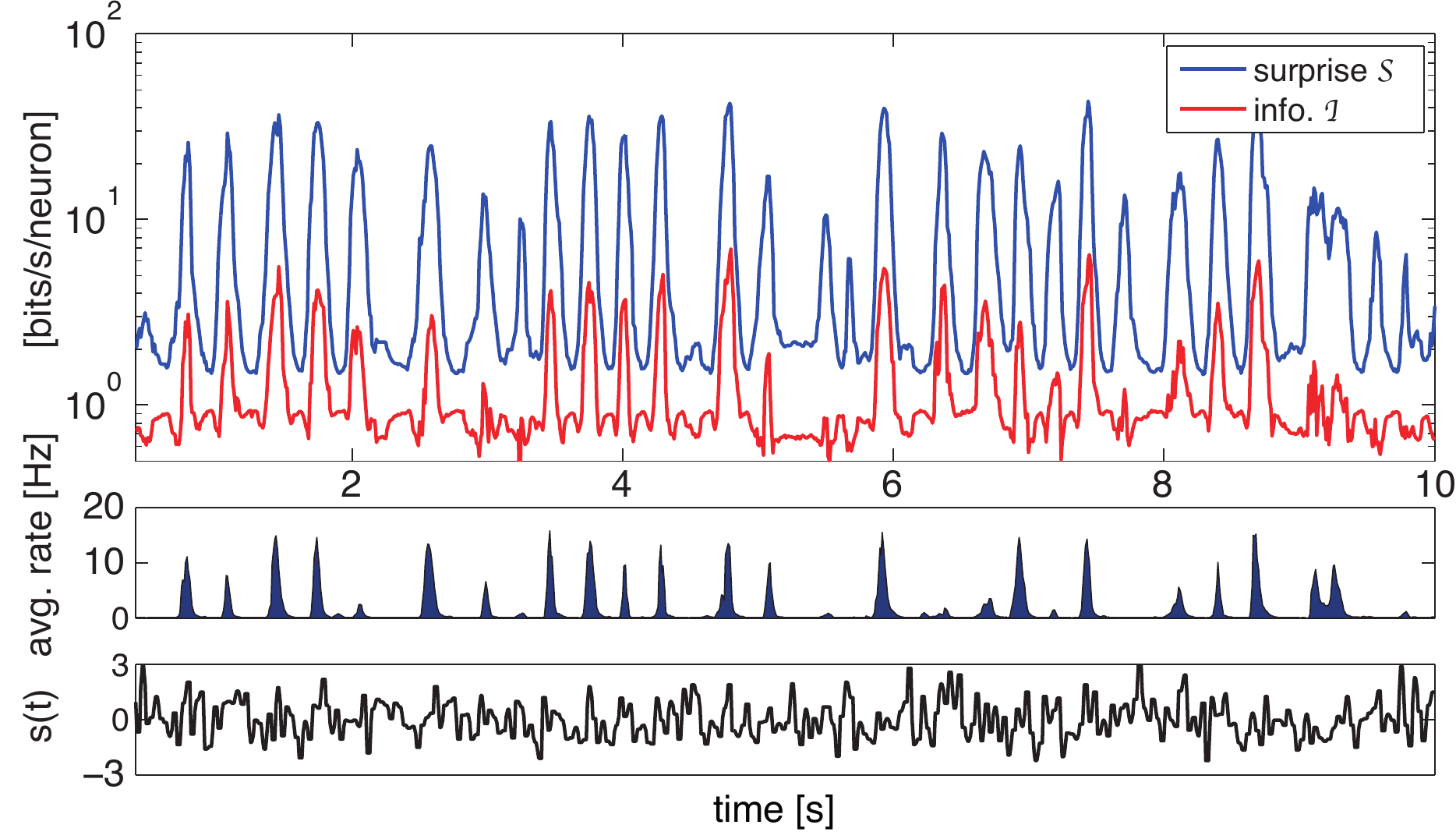} 
\caption{{\bf Surprise and information transmission estimated from the SDME model.} {\bf (a)}   Surprise rate (blue) is estimated from the static ME and SDME models assuming independence of codewords across time bins. The instantaneous information rate (red) is the difference between the surprise and the noise entropy rate, estimated from the SDME model (see text). The information transmission rate is the average of the instantaneous information across time.  {\bf (b)} Population firing rate as a function of time shows that bursts of spiking strongly correlate with the bursts of surprise and information transmission in the population.  {\bf (c)} The stimulus (normalized to zero mean and unit variance) is shown for reference as a function of time. }
\label{f8}
\end{figure}

The brain does not have direct access to the stimulus, but only receives codewords $\{x_i\}$, "drawn" from $P(\{x_i\})$, by the retina. At every moment in time, $-\log_2 P(\{x_i\})$ measures the {\em surprise}  about the output of the retina, and thus about the stimulus. We, as experimenters---but not the brain---have access to stimulus repeats and thus to $P(\{x_i\}|\mathbf{s}(t))$, so we can compute the average value of surprise (per unit time) at every instant $t$ in the stimulus:
\begin{equation}
\mathcal{S}(t) = -\frac{1}{\Delta t}\sum_{\{x_i\}} P(\{x_i\}|\mathbf{s}(t)) \log_2 P(\{x_i\}). \label{sur}
\end{equation}
This quantity can be expressed using the entropies and the learned parameters of our maximum entropy models, and is plotted as a function of time in Fig.~\ref{f8}. Since averaging across time is equal to averaging over the stimulus ensemble, we see from Eq.~(\ref{sur}) that $\langle \mathcal{S}(t)\rangle_t$ would have to be identically equal to $S[P(\{x\})]$ under the condition that $\langle P(\{x_i\}|\mathbf{s}(t))\rangle_t = P(\{x_i\})$ (marginalization). Since we build models for $P(\{x_i\})$ (static ME) and $P(\{x_i\}|\mathbf{s})$ (SDME) from data independently, they need not obey the marginalization condition exactly, but they will do so if they provide a good account of the data. Indeed, by using the static ME and SDME distributions in Eq.~(\ref{sur}) for surprise, we find that $\langle \mathcal{S}(t)\rangle_t\approx 740\e{bit/s}$, very close to the entropy rate $s$ of the total vocabulary and within our estimated 1\% error bars on entropy computation. 

To estimate the information transmission, we have to subtract the noise entropy rate from the output entropy rate $s$, as dictated by Eq.~(\ref{itrans}). The entropy of the SDME model is an upper bound on the noise entropy; since this is not a lower bound, we cannot put a strict  bound on the information transmission, but can nevertheless estimate it. Figure~\ref{f8} shows the ``instantaneous information'', $\mathcal{I}(t) = \mathcal{S}(t) - S[P^{SDME}(\{x_i\}|\mathbf{s}(t))]/\Delta t$, as a function of time; from Eq.~(\ref{itrans}), the mutual information rate is a time average of this quantity, $R=I(\{x_i\};\mathbf{s})/\Delta t=\langle\mathcal{I}(t)\rangle_t$. We find $R\approx 130\e{bit/s}$. This quantity can be compared to the total entropy rate of the stimulus itself (which must be higher than $R$), which in our case is $\approx 210\e{bit/s}$ (see Methods). While our estimates seem to indicate that a lot of vocabulary bandwidth (730 bit/s) is ``lost'' to noise (600 bit/s), the last comparison shows that the Gaussian FFF stimulus source itself is not very rich, so that the estimated information transmission  takes up more than half of the actual entropy rate of the source. 

Lastly, we asked how important is the inclusion of pairwise interactions, $\beta_{ij}$, into the SDME model, compared to a set of uncoupled LN models, when accounting for information transmission. We therefore estimated the noise entropy rate for a set of uncoupled LN models, $S[P^{LN}(\{x_i\}|\mathbf{s}(t))]/\Delta t$, which was found to be $\approx 770\e{bit/s}$, considerably higher than the noise entropy of the SDME model. Crucially, this noise entropy rate is larger than the total entropy rate $s$ estimated above, which is impossible for consistent models of the neural codebook and the vocabulary (since it would lead to negative information rates). This failure is a quantitative demonstration of the inability of the uncoupled LN models to reproduce the statistics of the population vocabulary, as shown in Fig.~\ref{f5}b, despite a seemingly small performance difference on the level of single cell PSTH prediction.

\section{Discussion}

We presented a modeling framework for stimulus encoding by large populations of neurons, which combines an individual neuronal receptive field model, with the ability to include pairwise interactions between neurons. The result is a stimulus-dependent maximum entropy (SDME) model, which is the most  parsimonious model of the population response to the stimulus that reproduces the linear-nonlinear (LN) aspect of single cells, as well as the correlation structure between neurons. In two limiting cases, the SDME model reduces to known models: if the single cell parameters $\alpha$ are static, SDME becomes the static maximum entropy model of the population vocabulary; if the couplings $\beta$ are 0, SDME becomes a set of uncoupled LN models. The framework is general, and could be easily applied to other neural systems. 

We applied this modeling framework to the salamander retina presented with Gaussian white noise stimuli, and found that the interactions between neurons play an important role in determining the detailed patterns of population response. In particular, the SDME model gave better prediction of PSTH of single cells, yielded orders of magnitude improvement in describing the population patterns, and captured significant aspects of noise correlations. The deviations between the SDME and the uncoupled LN model became significant for $>20$ cells, and tended to occur at ``interesting'' times in the stimulus, precisely when the neural population was not silent. 

The responses of the neural system in the maximum entropy framework are binary codewords of spiking and silence across the neural population. The choice of timescale over which these codewords are defined, here $\Delta t = 10\e{ms}$, is short enough such that multiple spikes are rarely observed in the same time bin, but long enough so that most of the strong spike-spike interactions (as well as fine temporal detail, such as spike-timing jitter) occur within a single bin. This allows us to view successive time bins as codewords, although some statistical dependence between them remains, possibly in the conditional SDME model (due to multiple timescales on which the neurons respond to stimuli), and certainly in the static ME model \cite{marre}. If we were to make the time scale much shorter, e.g. by an order of magnitude, we could make the conditional independence assumption of the responses given the stimuli \emph{and} previous spiking, which would lead us to GLM models \cite{pillow08} or nonequilibrium generalizations of Ising models \cite{roudi4}.  GLMs, in particular, are excellent generating models for precise spiking rasters, are easy to infer, and allow for asymmetric couplings between neurons. However, the inference in all these cases is tractable because there are no interactions between the spikes \emph{within} the same time bin (as there are in SDME). This necessitates the use of very short time bins and introduces  strong dependencies between successive time bins, making the interpretation of the discretized neural responses in terms of individual codewords difficult. For this reason, GLM and SDME are complementary approaches: the first allows for a temporally-detailed probabilistic description of a spiking  process, while the second gives an explicit expression for the probability distribution over codewords in longer temporal bins.

SDME allowed us to improve over LN models for salamander retinal ganglion cells both in terms of the PSTH prediction and, especially, in terms of population activity patterns. 
Interestingly, for parasol cells in the macaque retina  under flickering checkerboard stimulation,  the generalized linear model  did not yield firing rate improvement  relative to uncoupled LN models (but did improve higher order statistics of firing) \cite{pillow08}.
In both cases, however, the improvements reflect the role of dependencies among cells in encoding the stimulus, and their effect becomes apparent when we ask questions about information transmission by a neural population. Maximum entropy models can only put an upper bounds on the total entropy and the noise entropy of the neural code (and this statement remains true even if successive codewords are not independent), and as such cannot set a strict bound, but only give an estimate, for the information transmission. Nevertheless, ignoring the inter-neuron dependencies and using an uncoupled set of LN models predicts the total population responses so badly that the estimated noise entropy is higher than the upper bound on the total entropy, which is a clear impossibility, while the SDME model gives transmission rates that appear reasonable (and positive), amounting to about 60\% of the source entropy rate.
 
Tka\v{c}ik and colleagues \cite{tkacik10} have suggested that one can interpret $\beta_{ij}$ in an SDME model as  a prior over the activity patterns that the population would use to optimally encode the stimulus. For low noise level they argued that the prior should be minimal (and could help decorrelate the responses), as the population could faithfully encode the stimulus, whereas in the noisy regime, the prior should match the statistics of the sensory world and thus counteract the effects of noise. Similarly, Berkes and colleagues \cite{fiser} suggested a similar reason for the similarity of ongoing and induced activity patterns in the visual cortex. Our results  show that  interactions are necessary for capturing the network encoding, and implicitly reflect the existence of such a prior. The recovered interactions are strongly correlated with the interaction parameters  of a static, stimulus independent model over the distribution of patterns, making it possible for the brain (which only has access to the spikes, not the stimulus) to learn these values. Whether the interactions are matched to the statistics of the visual inputs as  suggested by Ref~\cite{tkacik10} is the focus of future work. In parallel, increasingly detailed statistical models of neural codes going beyond SDME (e.g. by including temporal dependencies as in Ref~\cite{cessac2}), and efforts to infer such models from experimental data, should focus our attention on population-level statistics and on  finding principled  information-theoretic measures for quantifying the code, like the surprise and instantaneous information suggested here. 
\section{Methods}
{\bf Electrophysiology.} Experiments were performed on the adult tiger salamander, \emph{Ambystoma tigrinum}. All experiments were in accordance with Ben-Gurion University of the Negev and government regulations.  Extracted retinas were placed with the ganglion cell layer facing a multielectrode array with
252 electrodes (Ayanda Biosystems, Switzerland), and superfused with oxygenated Ringer medium at room temperature. Extracellularly recorded signals were amplified (MultiChannel Systems, Germany) and digitized at 10k Samples/s, and spike-sorted using custom software written in MATLAB. 

{\bf Visual stimulation.}  Stimuli were projected onto
the retina from a CRT video monitor (ViewSonic G90fB) at a frame rate
of 60 Hz; each movie frame was presented twice, using standard optics. Full Field Flicker (FFF) stimuli were
generated by independently sampling spatially uniform gray levels (with a resolution of 8 bits) from a Gaussian distribution, with mean luminance of 147 lux and the standard deviation of 33 lux.  These data allow us to estimate the entropy rate of the source (as used in the main text), by multiplying the entropy of the luminance distribution with the refresh rate.
To estimate the cells' receptive fields, checkerboard stimulus was generated by selecting each
checker ($\sim 100 \e{\mu m}$ on the retina) randomly every 33 ms to be either black or white. To identify the RF centers, a two-dimensional Gaussian was fitted to the spatial profile of the response. The  movies were  gamma corrected for the computer monitor.  In all cases the visual stimulus entirely covered the retinal patch that was used for the experiment.

{\bf Inferring SDME from data.} 
The LN model for each neuron $i$ consists of the linear filter $\mathbf{k}_i$, and the nonlinear function $\mathcal{N}_i$, which is defined pointwise on a set of binned values for the generator signal, $g_i=\mathbf{k}_i\cdot \mathbf{s}$. We used binning into $K=20$ bins such that initially each bin contains roughly the same number of values for $g_i$, but subsequently the binning is adaptively adjusted (separately for each neuron) to be denser at higher values of $g_i$, where the firing rates are higher. We fitted LN models with varying number of $K$ bins, and have chosen $K=20$ when the performance of the LN models appeared to saturate \cite{granotmsc}. 

To find the parameters of the stimulus-dependent maximum entropy model ($\alpha_i(g_i),\beta_{ij}$), we retained the binning of the generator signal used for LN model construction.  Given trial values for the SDME parameters, we estimated the chosen expectation values (covariance matrix $C_{ij}$ in firing, and the firing rate conditional on $g_i$, $r_i(g_i)$) by Monte Carlo sampling from the trial distribution in Eq.~(\ref{eqs}); the learning step of the algorithm is computed by comparing the expectation values in the trial distribution and the empirical distribution (computed over the training half of the stimulus repeats). In detail, we used a gradient ascent algorithm, applying a combination of Gibbs sampling and
importance sampling in order to efficiently estimate the gradient, by using optimizations similar to those described in Ref.~\cite{Broderick+al_08}. Sampling was
carried out in parallel on a 16 node cluster with two 2.66GHz Intel Quad-Core Xeon
processors and 16GB of memory per node.  The calculation was terminated when the average error in firing rates and coincident firing rates reached below 1\% and 5\% respectively, which is within the experimental error.

To compute the single neuron PSTH and compare the distributions of codewords from the model to the empirical distribution, we used  Metropolis Monte Carlo sampling to draw codewords from the model distributions;  we drew 5000 independent samples (to draw uncorrelated configurations, a sample was recorded only after 100  ``spin-flip'' trials) for every timepoint, for a total of $5\E{6}$ samples; the same procedure was used also to draw from the uncoupled ($\beta=0$) models. To estimate the entropies of  high dimensional SDME distributions, we used the ``heat capacity integration'' method, detailed in Ref~\cite{preprint2}. Briefly, a maximum entropy model $P(\mathbf{x})=Z^{-1}\exp(-E(\mathbf{x}))$ (where $E$ is the Hamiltonian function determined by the choice of constrained operators and the conjugated parameters) is extended by introducing a new parameter $T$, much like the temperature in physics, so that $P_T(\mathbf{x})=Z_T^{-1}\exp(-E(\mathbf{x})/T)$. The entropy of the distribution is given by $S[P_{T=1}]=\int_0^1 C(T)/T dT$, where the heat capacity $C(T)=\sigma_E^2(T)/T^2$, and the variance in energy can be estimated at each $T$ by Monte Carlo sampling. In practice, we run a separate Monte Carlo sampling for a finely discretized interval of temperatures, $T\in[0,1]$, estimate $C(T)$ for each temperature, and numerically integrate to get the entropy $S$. We have previously shown that this procedure yields robust entropy estimates even for large numbers of neurons \cite{preprint,preprint2}.

\begin{acknowledgements}
{\bf Acknowledgements.} This work was supported by The Israel Science Foundation and the Human Frontiers Science Program.
\end{acknowledgements}
\end{document}